\documentclass[12pt]{article}
\usepackage{pdproc, epsfig} 
\usepackage{graphicx}
\usepackage{dcolumn}
\usepackage{amsmath}
\usepackage{enumerate}
\usepackage{subfigure}

  \def\be{\begin{equation}}
  \def\ee{\end{equation}}
  \def\bea{\begin{eqnarray}}
  \def\eea{\end{eqnarray}}

\begin{document}

\renewcommand{\thefootnote}{\alph{footnote}}
  
\title{
 CP VIOLATION IN NEUTRINO OSCILLATIONS \\ WITHOUT ANTINEUTRINOS:
                          ENERGY DEPENDENCE}

\author{Jose Bernabeu and Catalina Espinoza}

\address{ Departamento de Fisica Teorica and IFIC, CSIC-University 
of Valencia,\\ E-46100, Burjassot, Valencia, Spain \\
{\rm E-mail: jose.bernabeu@uv.es\\ m.catalina.espinoza@uv.es}}

\abstract{The next generation of long baseline neutrino oscillation
experiments will aim at determining the unknown mixing angle $\theta_{13}$,
the type of neutrino mass hierarchy and CP-violation. We discuss the
separation of these properties by means of the energy dependence of
the oscillation probability and we consider an hybrid setup which
combines the electron capture and the $\beta^+$ decay from the same
radioactive ion with the same boost. We study the sensitivity to the
mixing angle and the CP-phase, the CP discovery potential and the reach
to determine the type of neutrino mass hierarchy. The analysis is
performed for different boosts and baselines. We conclude that the
combination of the two decay channels, with different neutrino energies,
achieves remarkable results.}
   
\normalsize\baselineskip=15pt

\section{What is known, what is unknown}
In the past decade, atmospheric~\cite{SKatm,atm},
solar~\cite{sol,SKsolar,SNO}, reactor~\cite{CHOOZ,PaloVerde,KamLAND} 
and long-baseline accelerator~\cite{K2K,MINOS} neutrino  experiments
have provided compelling evidence for the phenomenon of neutrino
oscillations. This has reshaped our understanding of the properties of
elementary particles as it implies that neutrinos have mass and mix.
The combined data can be described by two mass squared differences,
$\Delta m_{31}^{2}$ and $\Delta m_{21}^{2}$, where $\Delta
m_{ji}^{2}=m_{j}^{2}-m_{i}^{2}$, whose current best fit values are
$|\Delta m_{31}^{2}|=2.4 \times 10^{-3} \ \mathrm{eV}^2$ and   
$\Delta m_{21}^{2}=7.65 \times 10^{-5} \ \mathrm{eV}^2$~\cite{STV08}. 
The two mixing angles $\theta_{12}$ and $\theta_{23}$ drive the solar
and KamLAND, and atmospheric and accelerator neutrino oscillations,
respectively, and are measured to be $\sin^2 \theta_{12} = 0.304$ and 
$\sin^2 \theta_{23} = 0.50$~\cite{STV08}. The third mixing angle,
$\theta_{13}$, is yet undetermined but is known to be small or
zero. With available data, $\theta_{13}$ is constrained to
be~\cite{STV08}
\begin{equation}
\sin^{2}\theta_{13} < 0.040 \: (0.056) \quad \mbox{at} \quad  2\sigma
\: (3\sigma) ~.
\end{equation}
It is interesting to note that very recently a first hint in favour
of $\theta_{13} \neq0$ has been found~\cite{Fogli:2008jx} in a
combined analysis of atmospheric, solar and long-baseline reactor
neutrino data, with: 
\begin{equation}\label{bound2}
\sin^2\theta_{13} = 0.016 \pm  0.010 \qquad \mbox{at 1$\sigma$} ~,
\end{equation}
implying a preference for $\theta_{13} > 0$ at 90\%~CL. The
combination of (\ref{bound2}) with other analyses is discussed~\cite{Ge:2008sj}
 by G.~L.~Fogli in the present Proceedings.

Although the experimental progress in neutrino physics over the last
decade has been conspicuous, many of the fundamental questions
surrounding neutrinos still need to be addressed. Understanding of the 
physics beyond the Standard Model responsible for neutrino masses and
mixing requires knowledge of the nature of neutrinos (whether Dirac
or Majorana particles), the neutrino mass ordering (normal or
inverted), the absolute neutrino mass scale, the value of the unknown  
mixing angle $\theta_{13}$, and whether CP-symmetry is violated in the
lepton sector. It will also be necessary to improve the precision on
the known parameters, in particular to measure any deviation from
maximal $\theta_{23}$ and, if so, to determine its octant.

Some of the issues above will be addressed by a future program of
neutrino oscillation experiments~\cite{ISS,Euronu}. For flavour oscillations, 
the unitary diagonalization of the neutrino mass matrix, assuming that the 
flavour mixing affects the three active light neutrinos only, is given by the
 PMNS matrix U which connects from mass eigenstate neutrinos to flavour 
eigenstate neutrinos

\be
\left[\begin{array}{ccc}
\nu_e \\
\nu_{\mu}\\
\nu_{\tau} \end{array}\right]
= U
\left[\begin{array}{ccc}
\nu_1 \\
\nu_2\\
\nu_3 \end{array}\right],
\ee

\bea
U =
\left[\begin{array}{ccc}
1  & 0  & 0 \\
0  & c_{23} & s_{23}  \\
0  & -s_{23} & c_{23}
\end{array}
\right]
\left[\begin{array}{ccc}
 c_{13} & 0  & s_{13}\, e^{-i\delta} \\
0  & 1 & 0  \\
-s_{13}\, e^{i\delta}  & 0 & c_{13}
\end{array}
\right]
\left[\begin{array}{ccc}
 c_{12} & s_{12} & 0 \\
-s_{12}  & c_{12} & 0  \\
0  & 0 & 1
\end{array}
\right]\,\,\,\,\,.
\label{Upartes}
\eea
$U$ is determined by 3 mixing angles and 1 CP-phase, even if neutrinos are
Majorana particles. The two additional Majorana phases would need a
$\Delta(L) = 2$ process to become observable, as in Neutrinoless Double
Beta Decay.

Nuclear reactors~\cite{freactors} and long baseline experiments using 
conventional beams~\cite{MINOS} will be the first to explore
$\theta_{13}$ below the current limit and maybe confirm the hint for
$\theta_{13} \neq 0$~\cite{Fogli:2008jx}. If $\theta_{13}$ is close to
the present bound imposed by the running and near future experiments,
the next generation of superbeams~\cite{T2K,newNOvA}, an extension of
a conventional beam with an upgrade in intensity and detector size,
and wide-band beams~\cite{Barger:2007jq} will probe CP-violation and,
for sufficiently long baseline, the neutrino mass hierarchy. For small
values of $\theta_{13}$ or, if  $\theta_{13}$ is large but a better
precision on the neutrino parameters needs to be achieved, the
community must turn to the novel concepts of the neutrino
factory~\cite{nufact,nufactlow} or beta-beam~\cite{zucchelli,mauro}.
Whereas conventional beams sourced from pion decays
have an intrinsic contamination of electron neutrino at the $\sim$ 1\%
level (owing to kaons in the beam), neutrino factories and beta-beams
will have clean sources from highly accelerated muons and ions,
respectively, producing a well-collimated beam. In a neutrino factory,
muons (antimuons) are produced, cooled and accelerated to a high boost
before being stored in a decay ring. The subsequent decay sources a
muon neutrino (muon antineutrino) and electron antineutrino (electron
neutrino) which are aimed at magnetised detectors located a very long
distance from the source. The use of magnetised detectors is necessary
to separate the `right muon' disappearance signal from the `wrong
muon' appearance signal, which is sensitive to matter effects and
CP-violation. A beta-beam will exploit accelerated ions that
$\beta$-decay sourcing a clean, collimated, electron neutrino beam.
Magnetised detectors will not be necessary in this case, the only
requirement being possession of good muon identification to detect the
appearance channels. Therefore, water \v{C}erenkov (WC), totally
active scintillator, liquid argon detectors and non-magnetised iron
calorimeters could be used, depending on the peak energy.

The determination of the oscillation parameters is severely affected
by degeneracies: the possibility that different sets of the unknown 
parameters $(\mbox{sgn}(\Delta
m_{31}^{2}),\delta, \theta_{13}, \theta_{23}$ octant) can provide an 
equally good fit to the probability for neutrino and antineutrino
oscillations, for fixed baselines and energy. Therefore, a high
precision measurement of the appearance probabilities is not
sufficient to discriminate the various allowed solutions. In order to
weaken or resolve this issue, various strategies have been put forward:
exploiting the energy dependence of the signal in the same experiment, 
using reactor neutrino experiments
with an intermediate baseline, combining different
long baseline experiments, adding the information on
$\theta_{13}$ from reactor experiments, or using
more than one baseline for the same
beam. In addition,  
$\theta_{13}$ controls the Earth matter effects in multi-GeV
atmospheric and accelerator neutrino oscillations, as well as in supernovae. 
These might provide useful information on the type
of neutrino mass hierarchy and $\theta_{13}$.

\section{CP-violation in neutrino oscillations}

The magnitude of the T-violating and CP-violating interference in neutrino oscillation
probabilities is directly proportional to $\sin\theta_{13}$~\cite{CPT,BB00}. 
CP-violation can be observed either by an Asymmetry between
neutrinos and antineutrinos and/or by Energy Dependence in the neutrino
channel. In the last case, the CP phase $\delta$ plays the role of a phase shift
in the interference pattern between the atmospheric and solar amplitudes
for the appearance oscillation probability. This result is a consequence
~\cite{BB00} of the assumptions of CPT-invariance and No Absorptive part in the
oscillation amplitude: the Hermitian character of the Hamiltonian
responsible of the time evolution says that the CP-odd$=$T-odd
probability $P (\nu_e \rightarrow \nu_{\mu}) - P (\bar{\nu_e} \rightarrow \bar{\nu_{\mu}})$  
is an odd function of time, i.e., an odd function of the baseline $L$. In vacuum neutrino 
oscillations for relativistic neutrinos, the oscillation phase depends
on the ratio $L/E$, then the CP-odd term becomes an odd function of the
energy $E$ for fixed $L$. With the same reasoning, the CP-even terms are
even functions of the energy $E$ in the oscillation probability. In 
this way, Energy Dependence in the appearance oscillation probability
is able to disentangle CP-even and CP-odd terms.

      The explicit expression for the suppressed appearance probability for neutrinos
in vacuum oscillations is given by
\bea
P(\nu_e \rightarrow \nu_\mu) & = &
s_{23}^2 \, \sin^2 2 \theta_{13} \, \sin^2 \left ( \frac{\Delta_{13} \, L}{2} \right ) +
c_{23}^2 \, \sin^2 2 \theta_{12} \, \sin^2 \left( \frac{ \Delta_{12} \, L}{2} \right ) \nonumber \\
& + & \tilde J \, \cos \left (\frac{ \Delta_{13} \, L}{2} + \delta \right ) \,
\frac{ \Delta_{12} \, L}{2} \sin \left ( \frac{  \Delta_{13} \, L}{2} \right ) \, ,
\label{vacexpand}
\eea
where $\Delta_{12} \equiv \Delta m^2_{21} /(2 E)$, $\Delta_{13}
\equiv \Delta m^2_{31} /(2E)$ and  $\tilde J \equiv c_{13} \, 
\sin 2 \theta_{12} \sin 2 \theta_{23} \sin 2 \theta_{13}$. 
This expression shows, in fact, that $|U(e3)|$ gives the strength of the probability,
whereas $\delta$ governs the interference pattern as a phase shift.
Furthermore, the separate atmospheric probability, the solar probability
and the CP-even term of the interference are even functions of $E/L$.
On the contrary, the CP-odd term of the interference between the
atmospheric and solar amplitudes is an odd function of $E/L$. This result
suggests the idea of disentangling $\delta$ from $|U(e3)|$ without a need
of comparing neutrino and antineutrino events, which have different
beam systematics and different cross sections in the detector: either monochromatic
neutrino beams with different boosts or a combination of channels
with different neutrino energies in the same boost are able of
separating the CP-violating phase.

Due to neutrino propagation through the Earth, matter effects
can "fake" CP-violation in the sense that the presence of matter affects
neutrino and antineutrino oscillations in a different way.
It is not easy to disentangle matter effects from CP-violation since
there is the so-called "mass hierarchy degeneracy", which swaps the
effect of matter for neutrino and antineutrino oscillation according to
the sign of $\Delta m^2_{31}$. The energy dependence of the $\nu_e \rightarrow \nu_{\mu}$
oscillation probability, in presence of matter effects, can be studied
from the expression~\cite{nf6,Akhmedov:2004ny}
\begin{equation}
\begin{array}{l}
\hspace{-8mm}P(\nu_e \rightarrow \nu_\mu, L)  = T_{atm} + T_{sol} + T_{int}\, ,  \\ \hspace{10mm} 
T_{atm} \simeq  
\sin^2 \theta_{23} \, \sin^2 {2 \theta_{13} } \left(
\frac{\Delta_{13}}{A- \Delta_{13}} \right)^2
\sin^2 \left( \frac{(A - \Delta_{13}) L}{2} \right), \\ \hspace{11mm}
T_{sol} = \cos^2 \theta_{23} \sin^2 {2 \theta_{12}} \left(
 \frac{\Delta_{12}}{A} \right)^2 \sin^2 \left( \frac{A L}{2} \right), \\ \hspace{11mm}
T_{int} = \tilde J \ \frac{\Delta_{12}}{A} \frac{\Delta_{13}}{A - \Delta_{13}} 
\sin \left(\frac{A L}{2} \right) \sin \left( \frac{(A -
  \Delta_{13}) L}{2} \right)
 \cos \left(\frac{\Delta_{13} L}{2}+ \delta \right),
\end{array}
\label{eq:probappr}
\end{equation}
where $A \equiv \sqrt{2} G_{F} \bar{n}_e(L)$ and $\bar{n}_e(L)= 1/L \int_{0}^{L} n_e(L') dL'$ is
the average electron number density. As seen, the energy dependence induced by the presence of 
$A\ne0$ is different in $T_{atm}$, $T_{sol}$ and in the interference $T_{int}$ and, in fact, 
different from the energy dependence associated with the CP-even versus the CP-odd separation. 
On the other hand, the mass hierarchy degeneracy in vacuum
is now removed  because $T_{atm}$ and $T_{int}$
are now changing under the change of sign of $\Delta m^2_{31}$, although $T_{sol}$ remains the same. 
All in all, we observe the virtues of studying the neutrino appearance probability
as a function of the neutrino energy.

From the current discovery phase of $\theta_{13}$, next generation
experiments will hence aim at precision measurements of the $\nu_e \rightarrow \nu_{\mu}$
oscillation probability. This will require large underground detectors 
coupled to more intense and pure neutrino beams. These aspects are being
studied within the LAGUNA and EURONu design studies. The knowledge
on the possible values of $\theta_{13}$ is a necessary input to best optimize
the search for CP-violation in the leptonic sector.

The energy dependence of the signal is
typically used to extract information on the mass hierarchy and
CP-violation. Matter effects increase with baseline and energy
suggesting that setups with baselines L$>$ 600 km are necessary~
\cite{mantle,core,atmmatter2}
for the determination of the type of neutrino mass ordering. In beta-beam 
experiments, such strategies would make use of a proposed upgrade of the CERN Super
Proton Synchrotron (SPS) which would equip the accelerator with fast 
superconducting magnets allowing high boosts and fast ramps. The
latter are important to reduce the loss of ions through decay in the
acceleration stage. A sister approach to the beta-beam is to use the
neutrinos sourced from ions that decay mainly through electron
capture (EC)~\cite{EC1,EC2,Sato,RS06}. If the
electron capture decay is dominated by a single channel, then a 
monoenergetic electron neutrino beam can be produced this way. In this
case, all the beam intensity can be concentrated at the appropriate
energy to get the best sensitivity to the oscillation parameters. In
order to disentangle the CP violating phase with neutrino beams only,
one makes use of the different energy dependence of the CP-even and
CP-odd terms in the appearance probability~\cite{BB00}. Electron
capture competes with $\beta^{+}$-decay when the $Q_{\rm EC}$-value $> 
2m_{e}$, $m_{e}$ being the electron mass. With the ions identified
in~\cite{EC1}, the use of an upgraded SPS or the
Tevatron~\footnote{Note that the present Tevatron configuration does 
  not ramp fast enough, resulting in a high loss of ions, so this
  might not be a very realistic experimental setup, at least in the
  present configuration.} is necessary to source baselines in excess
of CERN-Frejus (130 km). We will discuss here also a hybrid 
\cite{Bernabeu:2009np} of these two approaches. By selecting a 
nuclide with $Q_{\rm EC}\sim 4$~MeV, we can make use of neutrinos from
an electron capture spike and $\beta^{+}$ continuous spectrum
simultaneously. Assuming a detector with low energy threshold, the use
of such ions allows one to exploit the information from the first and
second oscillation maxima with a single beam. The use
of the hybrid approach makes it possible to use a
monochromatic beam at higher energies and a beta-beam at lower
energies. The need for good neutrino energy resolution at the higher
energies will therefore be less crucial than for high-$\gamma$
beta-beam scenarios.

\section{A combined beta-beam and electron capture neutrino experiment}

In this section we discuss the idea \cite{Bernabeu:2009np} of the 
beta-beam and electron capture hybrid approach. We present the spectra 
of the two branches, their ratio and consider two nuclides which have 
desirable properties. 

The beta-beam is a proposal, originally put forward by
P.~Zucchelli~\cite{zucchelli}, to accelerate and then store
$\beta$-emitting ions, which subsequently decay to produce a well
collimated, uncontaminated, electron neutrino (or antineutrino)
beam. The high luminosities required to achieve a useful physics reach
point towards ions with small charge to minimise space charge
and half-lives $\sim$ 1~second to reduce ion losses during the
acceleration stage whilst maintaining a large number of useful decays
per year. The most promising candidate ions are $^{18}$Ne and $^{8}$B
for neutrinos, and $^{6}$He and $^{8}$Li for antineutrinos. A variant
on the beta-beam idea is the use of electron capture to produce
monoenergetic neutrino beams. Electron capture is the process in which
an atomic electron is captured by a bound proton of the ion $A(Z,N)$
leading to a nuclear state of the same atomic number $A$, but with the
exchange of the proton by a neutron and the emission of an electron
neutrino, 
\begin{equation}
A(Z,N) + e^- \rightarrow A(Z-1,N+1) + \nu_e ~.
\end{equation}
The idea of using this process in neutrino experiments was
independently discussed in Refs.~\cite{EC1,Sato}. In Ref.~\cite{RS06},
ions with low $Q_{\rm EC}$-value and long half-life, such as
$^{110}$Sn, were proposed to be accelerated to very high boosts with
the LHC. Baselines of 250~km and 600~km were considered with the
spectral information coming from the position of the events in the
detector. Sensitivities comparable to a Neutrino Factory were obtained
for a single boost. However, in order for electron capture machines to
become operational, nuclei with shorter half-life are required. The
recent discovery of nuclei far from the stability line with
kinematically accessible super-allowed spin-isospin transitions to 
giant Gamow-Teller resonances (see, for example, Ref.~\cite{RG09})
opens up such a possibility. The rare-Earth nuclei above $^{146}$Gd
have a short enough half-life to allow electron capture processes in
the decay ring, in contrast to fully-stripped long-lived
ions~\cite{Sato,RS06}. This was the scenario put forward in
Ref.~\cite{EC1} where the use of short-lived ions with $Q_{\rm
  EC}$-values around 1-4 MeV was proposed. Machines such as the SPS,
an upgraded SPS and the Tevatron could then be used for the
acceleration. The ion $^{150}$Dy, with $Q_{\rm EC}$-value 1.4~MeV, was
investigated for the CERN-Frejus (130 km) and CERN-Canfranc (650 km)
baselines and different boost factors. It was found to have very good
physics reach~\cite{EC1,EC2}. Owing to the monochromatic nature of the
beam, multiple boosts are necessary to resolve the intrinsic
degeneracy in this case.

In the following, we demonstrate how the flux for the electron
capture/beta-beam can be built up by discussing them separately and
comparing branching ratios. Let the mass difference between the parent
and the daughter atoms, $\Delta M_A^{\beta^+} = M_A(Z,N) -
M_A(Z-1,N+1)$, include the mass and the binding energy of an atomic
electron as well. For electron capture, the maximum kinetic energy
release is thus given by $Q_{\rm EC} =  \Delta M_A^{\beta^+}$. 
For $\beta^+$-decay, however, the final atom has an excess electron
since a positron is produced. The maximum kinetic energy release is
thus given by $Q_{\beta^+} = \Delta M_A^{\beta^+} - 2 \, m_{e}$.
Clearly for $(\Delta M_A^{\beta^+} = ) \, Q_{\rm EC} <  2m_{e}$,
electron capture is the only allowed process for a proton-rich
nucleus. For $Q_{\rm EC} > 2m_{e}$, electron capture and positron
emission compete, their branching ratios dependent on $Q_{\rm EC}$. If
decay through $\alpha$ emission is also allowed, it is important that
this has a relatively\, low\,\, $Q$-value so as not to be the dominant
channel~\footnote{The $\alpha$ decay branching ratio is strongly
  dependent on the $Q_{\rm EC}$-value. For low $Q_{\rm EC}$, the
  $\alpha$ decay probability is sufficiently long as to allow the weak
  decay modes to be the main channels.}. For a number of useful ion
decays per year $N_{\rm ions}$, the electron capture neutrino flux is
given by~\cite{EC1,EC2}
\begin{equation}
\frac{d\Phi^{\rm lab}_{\rm EC}}{d\Omega
  dE_\nu}=\frac{\Gamma}{\Gamma_{\rm tot}}\,\frac{N_{\rm ions}}{\pi
  L^{2}}\,\gamma^{2}\,\delta(E_\nu-2\gamma E_0^{\rm EC})
\end{equation}
for each decay channel. Here, $L$ is the baseline, $\gamma$ is the
Lorentz boost, $E_0^{\rm EC}$ ($= Q_{\rm EC})$ is the neutrino energy
in the ion rest frame and $E_\nu$ is the neutrino energy in the lab
frame.

The flux for the $\beta$-spectrum is found in the usual way. In the
rest frame of the ion, the electron neutrino flux is proportional to 
\begin{equation}
\frac{d\Phi^{\rm rf}_\beta}{d\cos\theta dE_{\rm rf}} \sim
E_{\rm rf}^2 (E_0^\beta-E_{\rm rf})\sqrt{(E_{\rm rf} - E_0^\beta)^{2}
  - m_{e}^{2}}
~. 
\label{E:flux}
\end{equation}
Here, $E_0^\beta$ ($= Q_{\beta^+} + m_e = Q_{\rm EC} - m_e$) is the
total end-point energy of the decay. The neutrino flux per solid angle
at the detector located at distance $L$ from the source after boost
$\gamma$ is~\cite{betabeamhigh}  
\begin{equation}
\left.\frac{d\Phi^{\rm lab}_\beta}{d\Omega dy}\right|_{\theta\simeq 0}
\simeq \frac{N_{\rm ions}}{\pi  L^{2}}
\frac{\gamma^{2}}{g(y_{e})}y^{2}(1-y)\sqrt{(1-y)^{2}-y_{e}^{2}} ~,
\label{E:Bflux}
\end{equation}
where $0 \leq y=\frac{E_{\nu}}{2\gamma E_0^\beta}\leq 1-y_{e}$,
$y_{e}=m_{e}/E_0^\beta$, and 
\begin{equation}
g(y_{e})\equiv \frac{1}{60}\left\{
\sqrt{1-y_{e}^{2}}(2-9y_{e}^{2}-8y_{e}^{4})+15y_{e}^{4}
\log\left[\frac{y_{e}}{1-\sqrt{1-y_{e}^{2}}}\right] \right\}. 
\end{equation}
Similarly to the case of electron capture, a neutrino with energy
$E_{\rm rf}$ in the rest frame will have a corresponding energy
$E_{\nu} = 2\gamma E_{\rm rf}$ in the laboratory frame along the
$\theta =0^{\circ}$ axis.  

All the known nuclear structure information on the $A=148$ and $A=156$
nuclides has been reviewed in Ref.~\cite{Bhat} and Ref.~\cite{Reich},
respectively, where the information obtained in various reaction and
decay experiments is presented, together with adopted level schemes.
Currently, a systematic study of electron capture decays in the region
of $^{146}$Gd, relevant for monoenergetic neutrino beams, is being
carried out~\cite{Ybcom}. Here, we consider two nuclides,
$^{156}_{70}$Yb and $^{148m}_{65}$Tb, that decay through electron
capture and $\beta^{+}$-decay with similar branching ratios whose
lifetimes are not too long or too short. Their decays are summarised
in Tables~\ref{T:decay_Yb} and~\ref{T:decay_Tb}. Ytterbium is a
nuclide $^{156}_{70}$Yb with spin-parity $0^+$, which decays 90\% via
electron capture plus $\beta^+$-decay~\cite{Reich}, with 38\% via
electron capture and 52\% via $\beta^+$-decay~\cite{Ybcom}. The
remaining 10\% goes into $\alpha$-particles and a different final
state. This relatively small branching ratio into $\alpha$'s helps the
nuclide to have a short enough half-life, 26.1~seconds. It is
important to note that this electron capture-$\beta^+$-decay
transition has only one possible daughter state with spin-parity
$1^+$, i.e., it is a Gamow-Teller transition into an excited state of
Thulium, $^{156}_{69}$Tm$^*$. The transition $Q_{\rm EC}$-value
is~\footnote{$Q_{\rm EC}$-values are typically calculated between
  ground states unless stated otherwise.} $Q_{\rm EC}$-value = 3.58
MeV. However, the excitation energy of the final nuclear state
(0.12~MeV) needs to be taken into account and thus, the effective
$Q_{\rm EC}$-value (difference in the total kinetic energies of the
system after and before the decay) is 3.46~MeV~\cite{Reich}. The
electron capture energy of $\sim 4$ MeV is well suited to the
intermediate-baselines of Europe and the USA with the available
technology, or those available with future upgrades. On the other 
hand, the $^{148m}_{65}$Tb isomer with spin-parity $9^+$ has a $Q_{\rm EC}$-value 
of 5.77~MeV~\cite{Bhat,Tb}. Although the decay to the
ground state of $^{148}_{64}$Gd is highly forbidden, the presence of a
Gamow-Teller resonance allows the decay into an excited state with
effective $Q$-value 3.07~MeV~\cite{GSI}. This nuclide is longer lived  
than $^{156}_{70}$Yb (with a half-life of 2.2 minutes) and will require
slightly higher boosts. It is still well suited to intermediate
baselines. However, the dominance of the electron capture over the
$\beta^{+}$-decay channel makes this nuclide less desirable. The count
rate will be dominated by the single energy of the electron capture
which provides insufficient information to obtain the good
sensitivities aspired to by future long baseline experiments. It was
shown in Refs.~\cite{EC1,EC2} that two runs with different boosts are
necessary for an exclusive or dominant electron capture channel to
break the intrinsic degeneracy and achieve good CP-violation
discovery. Hence, in what follows we will study this hybrid approach
focusing on $^{156}$Yb.

\begin{table}
\begin{center}
\begin{tabular}{|c|c|c|c|}
\hline
\quad Decay \quad & \quad Daughter \quad & \quad Neutrino energy (MeV)\quad & \quad BR \quad \\  
&&& \\
\hline
\hline
$\beta^{+}$ & $^{156}_{69}$Tm$^{*}$ & 2.44 {\rm (endpoint)} & 52\% \\
EC & $^{156}_{69}$Tm$^{*}$ & 3.46 \hspace{1.77cm} & 38\%\\
$\alpha$ & $^{152}_{68}$Er &  \hspace{1.77cm} & 10\% \\
\hline
\end{tabular}
\end{center}
\caption{Decay summary for $^{156}_{70}$Yb. The $Q_{\rm EC}$-value for
the transition between ground states is 3.58 MeV and taking into
 account the excitation energy of the final nuclear state (0.12~MeV),
 the effective $Q^{\rm eff}_{\rm EC}$-value is 3.46~MeV.}
\label{T:decay_Yb}
\end{table}

\begin{table}
\begin{center}
\begin{tabular}{|c|c|c|c|}
\hline
\quad Decay \quad & \quad Daughter \quad & \quad Neutrino energy (MeV)
\quad & \quad BR \quad \\ 
& & & \\
\hline
\hline
$\beta^{+}$ & $^{148}_{64}$Gd$^{*}$ & 2.05 {\rm (endpoint)} & 32\% \\
EC & $^{148}_{64}$Gd$^{*}$ & 3.07 \hspace{1.77cm} & 68\%\\
\hline
\end{tabular}
\end{center}
\caption{Decay summary for $^{148\, m}_{65}$Tb. The $Q_{\rm EC}$-value
  for the transition between ground states is 5.77~MeV and the
  effective $Q^{\rm eff}_{\rm EC}$-value to the excited state is
 3.07~MeV.} 
\label{T:decay_Tb} 
\end{table}

\section{Choice of the boost and baseline}
We consider the use of a neutrino beam sourced from
boosted $^{156}$Yb ions directed along a single baseline. As described
above, both the electron capture and $\beta^{+}$-decay channels are to
an excited state of $^{156}$Tm with a $Q_{\rm EC}$-value of
3.46~MeV. In order to fully exploit the electron capture decay mode,
the nuclides cannot be fully stripped; at least 16 electrons being
left on the ion~\cite{Mats}. The maximum boost, $\gamma_{\rm max}$,
available is thus
\begin{equation}
\gamma_{\rm max}=\frac{E_{\rm acc}}{m_{p}}\frac{Z-16}{A} ~,
\end{equation}
where $m_{p}$ is the mass of the proton and $E_{acc}$ is the maximum
energy accessible with the accelerator. Current and future accelerator
facilities would be an ideal production environment. In this analysis,
we consider the maximum boosts available from the current SPS and
upgraded SPS (see Table~\ref{T:maxboosts}) for the following baselines:
\begin{figure}
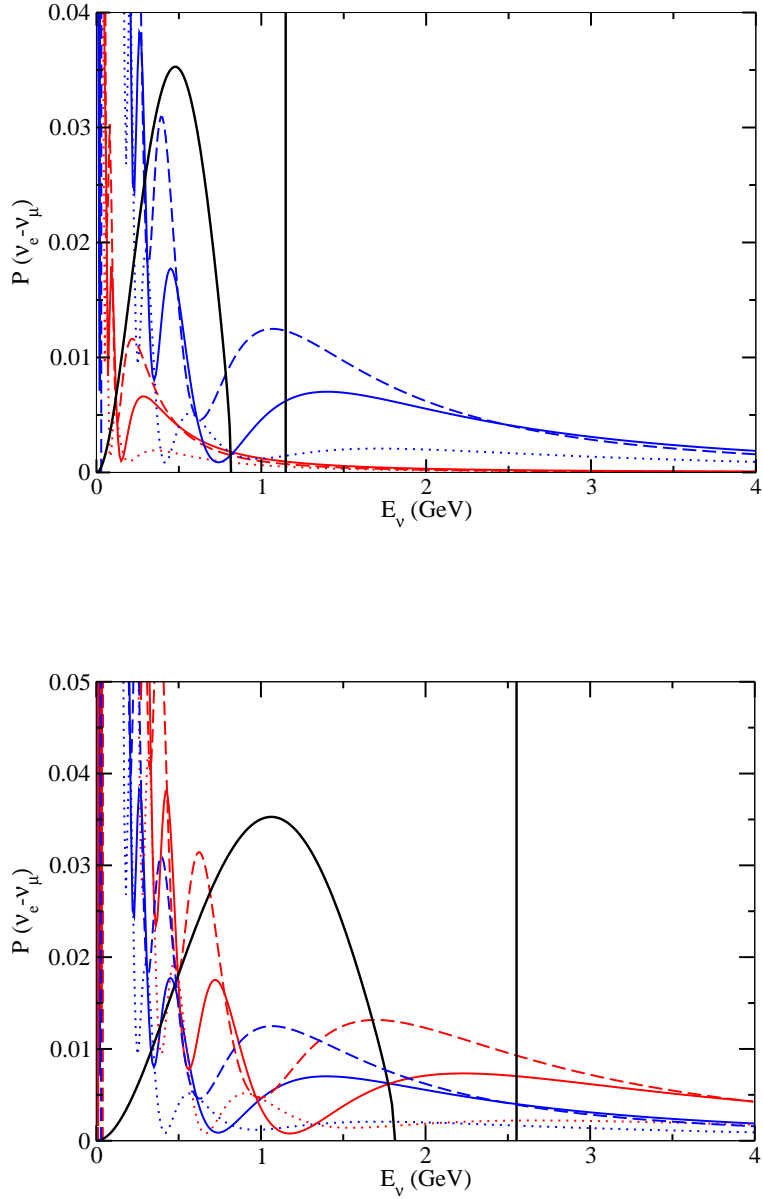

\begin{center}
\subfigure{\includegraphics[width=10cm,height=7cm]{g166.eps}}\\
\vspace{1.5cm}
\subfigure{\includegraphics[width=10cm,height=7cm]{g369.eps}}
\end{center}
\caption{{\it Top panel}: $\nu_{e}\rightarrow \nu_{\mu}$ appearance
  probabilities for the CERN-Frejus (130 km) and CERN-Canfranc (650 km)
  baselines. The unoscillated $\nu_{e}$ flux in the laboratory frame
  is shown for $^{156}$Yb given a boost $\gamma=166$ in arbitrary
  units. {\it Bottom panel}: $\nu_{e}\rightarrow \nu_{\mu}$ appearance
  probabilities for the CERN-Canfranc (650 km) and CERN-Boulby (1050
  km) baselines. The flux from a boost $\gamma=369$ is shown in
  arbitrary units. In both cases, the blue lines correspond to
  CERN-Canfranc; the red being CERN-Frejus (top panel) and CERN-Boubly
  (bottom panel). The solid lines correspond to $\delta=0^{\circ}$,
  dashed $\delta=90^{\circ}$ and dotted   $\delta=-90^{\circ}$. The
  value $\sin^2 2\theta_{13} = 0.01$ was taken for all curves.} 
\label{F:ECprobs}
\end{figure}

\begin{enumerate}
\item \textbf{Boost $\gamma=166$ with current SPS}
\begin{itemize}
\item CERN-Frejus (130 km)
\item CERN-Canfranc (650 km)
\end{itemize}
\item \textbf{Boost $\gamma=369$ with an upgraded SPS}
\begin{itemize}
\item CERN-Canfranc (650 km)
\item CERN-Boulby (1050 km)
\end{itemize}
\end{enumerate} 
\begin{table}
\begin{center}
\begin{tabular}{|c|c|c|c|}
\hline
\; Machine\; &\; $\gamma_{\rm max} $\; &\; $2 \gamma_{\rm max} Q^{\rm eff}_{\rm EC}$(GeV)\; &\;
 $2 \gamma_{\rm max} Q^{\rm eff}_{\beta^{+}} $ (GeV)\; \\
\hline
\hline
SPS & 166 & 1.15 & 0.81\\
\quad Upgraded SPS \quad & 369 & 2.55 & 1.80 \\
\hline
\end{tabular}
\end{center}
\caption{Maximum boosts and neutrino endpoint energies for $^{156}$Yb
  available for the current SPS setup and a proposed 1 GeV upgraded
  SPS.}
\label{T:maxboosts}
\end{table}
With the current magnetic rigidity of the SPS,  the electron capture
spike can be placed on first oscillation for the CERN-Canfranc
baseline (650 km) with the beta-beam spectrum peaking around the
second oscillation maximum (see Fig.~\ref{F:ECprobs}). For the upgraded SPS, 
with proton energy at 1~TeV, the first oscillation maximum and most of the 
second oscillation are covered for the CERN-Boulby baseline (1050 km).

\newpage

\section{Results}
We will compare the physics reach for different experimental
Setups, defined by the following:

\begin{enumerate}
\item \textbf{50~kton detector (LAr or TASD) with $2 \times
  10^{18}$~ions/yr}
\begin{itemize}
\item Setup I: CERN-Frejus (130~km) and $\gamma = 166$
\item Setup II: CERN-Canfranc (650 km) and $\gamma = 166$
\item Setup III: CERN-Canfranc (650~km) and $\gamma = 369$
\item Setup IV: CERN-Boulby (1050~km) and $\gamma = 369$ 
\end{itemize}
\item \textbf{0.5~Mton water-\v{C}erenkov detector with
  $2 \times 10^{18}$~ions/yr} 
\begin{itemize}
\item Setup III-WC: CERN-Canfranc (650 km) and $\gamma = 369$
\item Setup IV-WC: CERN-Boulby (1050 km) and $\gamma = 369$
\end{itemize}
\end{enumerate} 

\noindent I and II correspond to present SPS energies for the boost,
whereas III and IV need an upgraded SPS with proton energy 1 TeV.
The difference between III(IV) and III(IV)-WC is in the detector.
In the first case, we consider a 50 kton detector, like LiAr or TASD,
with energy reconstruction and neutrino spectral information. In the
second case, a $0.5$ Mton Water Cerenkov detector with neutrino energy
from QE events only plus inelastic events in a single bin, with $70\%$
efficiency. Whereas in the first case we use then the total cross section
 at each energy, for the WC detector we use the total cross section at the 
EC spike, the QE cross section at each energy bin and the inelastic cross 
section for a single bin in the  $\beta^+$ spectrum.

       The separation between the energy of the EC spike and the end
point energy of the beta-spectrum is possible: if $E_{\nu}({\rm QE}) > 2
\gamma E_{0}^{\beta}$, since $E_{\nu}^{{\rm true}} \geq
E_{\nu}({\rm QE})$, the event has to be attributed
to the EC flux and, hence, it is not necessary to reconstruct the true
neutrino energy.

\subsection{{\bf Comparing baselines}}

\begin{figure}
\begin{center}
\hspace{-1.5cm}
\subfigure{\includegraphics[width=8.9cm]{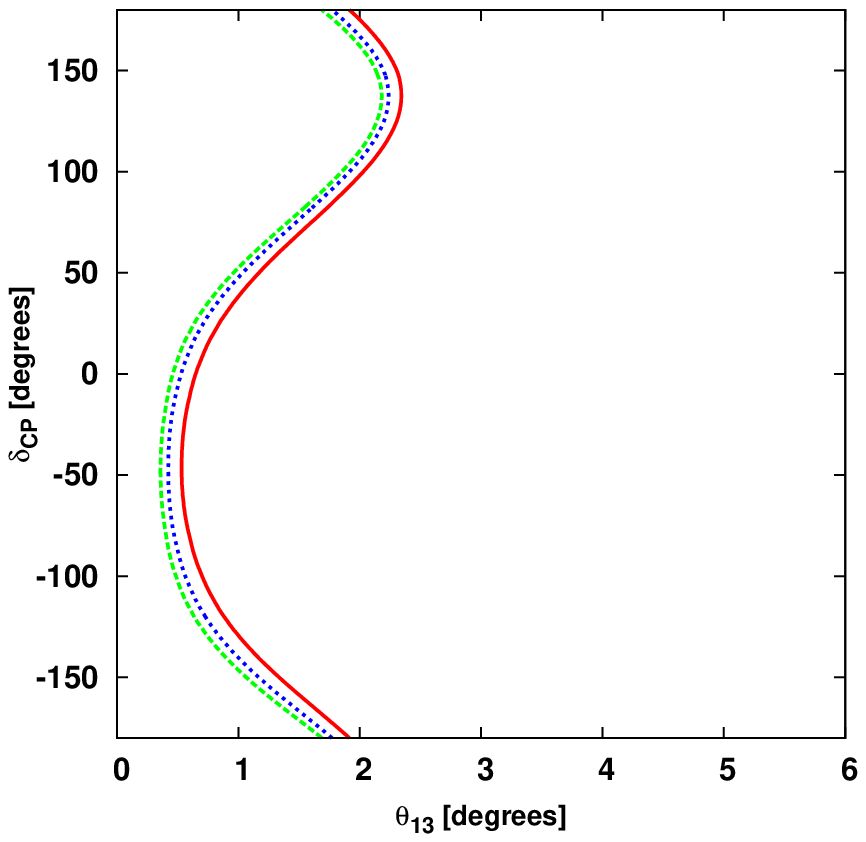}} \hspace{-1.5cm}
\subfigure{\includegraphics[width=8.9cm]{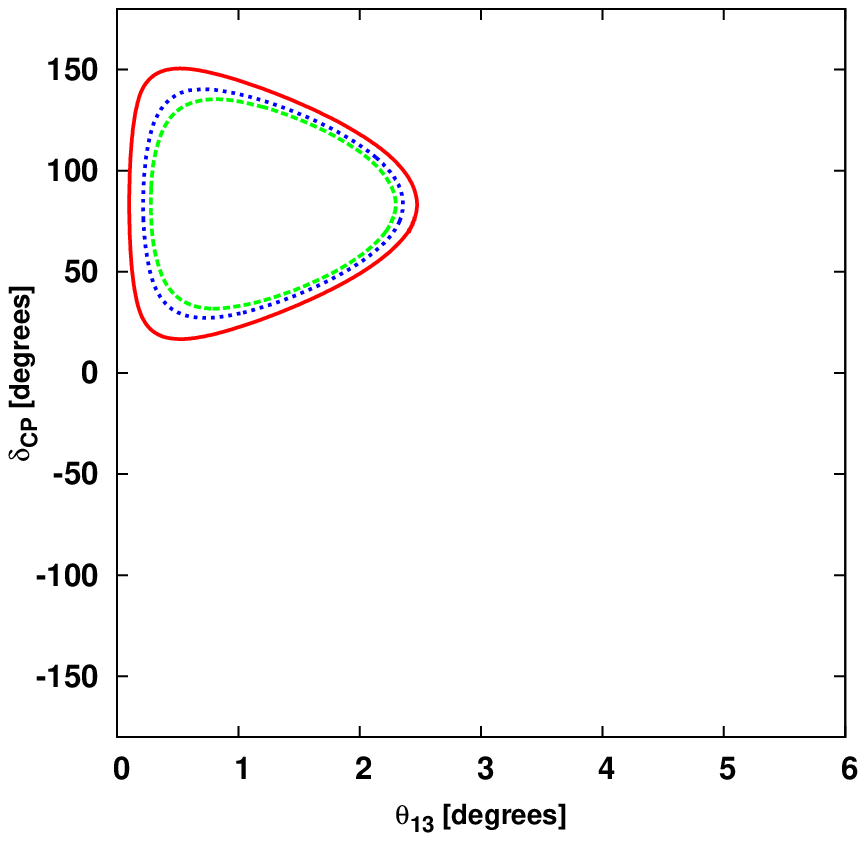}}
\end{center}
\vspace{-2mm}
\caption{90\%, 95\% and 99\% CL contours for setup I (left panel) and
  setup II (right panel). The parameters $\theta_{13}=1^{\circ}$ and
  $\delta=90^{\circ}$ have been taken assuming normal mass ordering and
  $\theta_{23}=45^{\circ}$.}
\label{Fi:C166}
\end{figure}

For this comparison, we take the combined beta-beam and EC fluxes
with $\theta_{13}= 1^o$ and $\delta=90^o$ for Setups I and II, with different
baselines. The result is shown in Fig.~\ref{Fi:C166}, where we see that the
 longer baseline (650 km) is necessary in order to separate the mixing parameter
and the CP phase. Violation of CP can be established, only in the last
case, for a limited range of values of the parameters. 

In fact, the beta-beam channel contributes very little to
the overall sensitivity of the setup. This is due to the energy dependence of the flux. 
The smaller flux, combined with the smaller cross section at the energies centred on 
second oscillation maximum, supplies a scarce count rate. The bulk of
the sensitivity is due, in this case, to the electron capture channel,
placed on first oscillation maximum, as seen in Fig.~\ref{F:ECprobs}. 

\subsection{{\bf Comparing boosts at the same baseline}}
We are going to compare the results for the setups II and III.
In going  from $\gamma = 166$ to $\gamma = 369$, the electron capture beam is
placed in the tail of first oscillation maximum. This gives the
beta-beam coverage of the second oscillation maximum and substantial
portions of the first oscillation maximum. Then the roles of electron
capture and beta-beam are reversed. The beta-beam now contributes much
more to the sensitivity as it provides substantial information from 
the first oscillation maximum and a much higher count rate from the
second oscillation maximum.

\begin{figure}
\begin{center}
\hspace{-1.5cm}
\subfigure{\includegraphics[width=8.9cm]{PatSII-1-90-EC+BB.eps}} \hspace{-1.5cm}
\subfigure{\includegraphics[width=8.9cm]{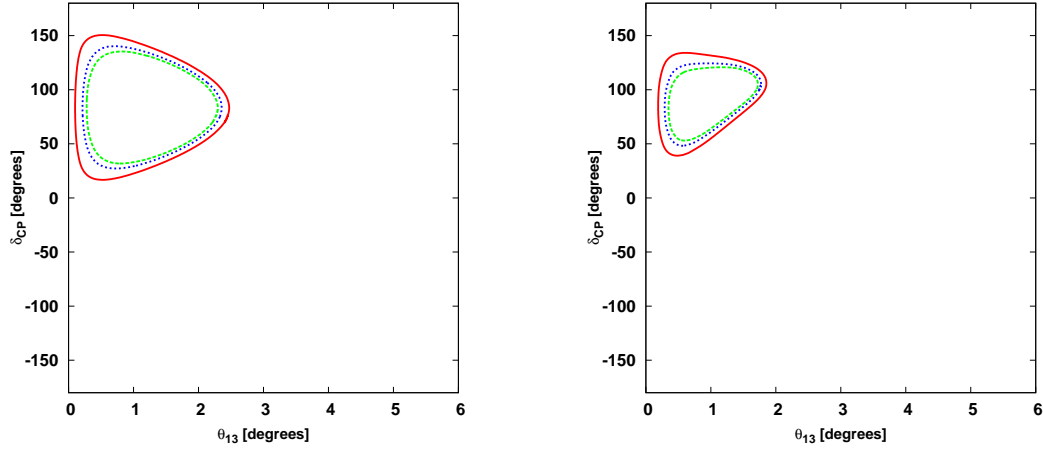}}
\end{center}
\vspace{-2mm}
\caption{90\%, 95\% and 99\% CL contours for setup II (left panel) and
  setup III (right panel). The parameters $\theta_{13}=1^{\circ}$ and
  $\delta=90^{\circ}$ have been taken assuming normal mass ordering and
  $\theta_{23}=45^{\circ}$.}
\label{Fi:C166-1}
\end{figure}

In Fig.~\ref{Fi:C166-1} we show the 90\%, 95\% and 99\% CL contours for setups
II and III with mixing $\theta_{13} = 1^o$ and $\delta = 90^o$, using the
combination of beta-beam and EC fluxes. We conclude that the sensitivity
is better with the upgraded SPS energy.

\subsection{{\bf The virtues of combining energies from beta-beam and EC}}
\begin{figure}
\begin{center}
\hspace{-1.5cm}
\subfigure{\includegraphics[width=6.37cm]{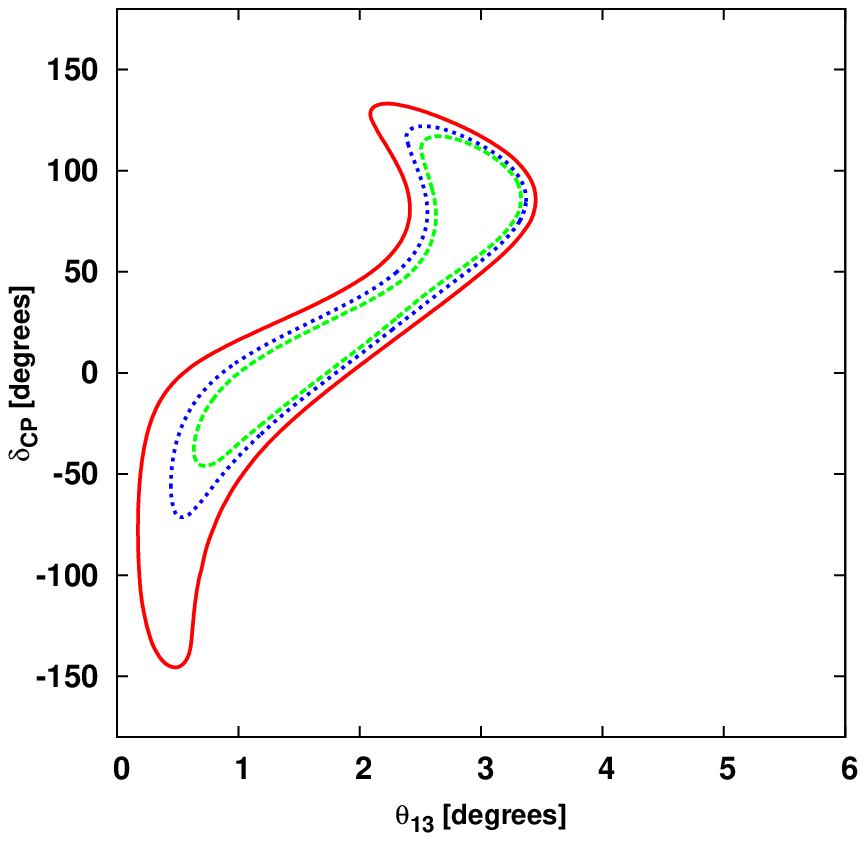}}
\hspace{-1.5cm}
\subfigure{\includegraphics[width=6.37cm]{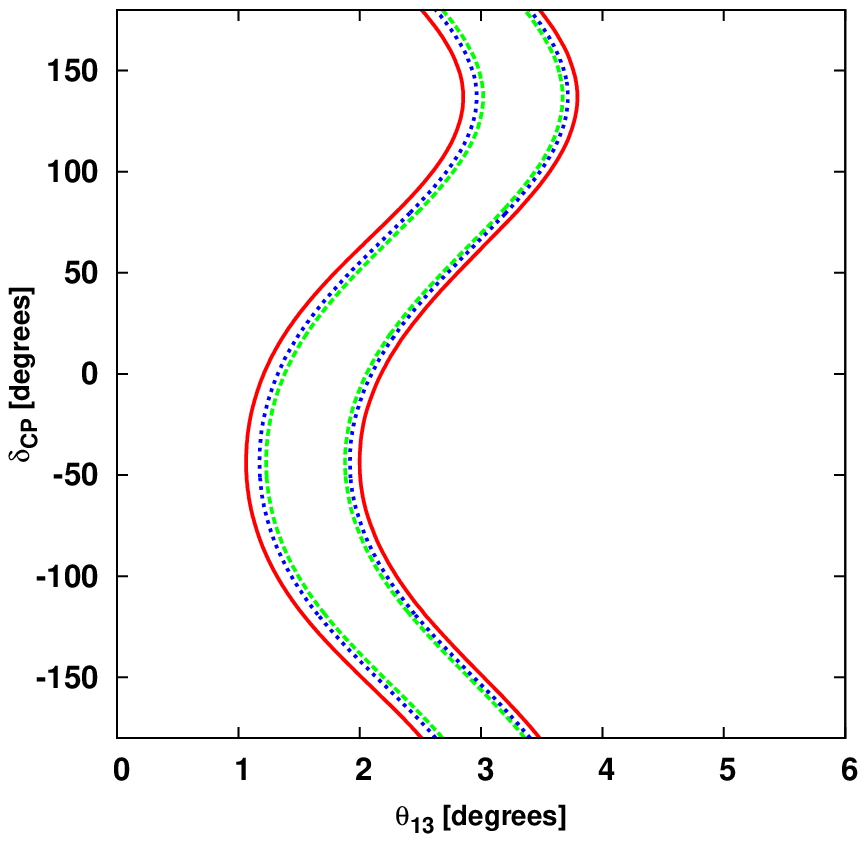}}
\hspace{-1.5cm}
\subfigure{\includegraphics[width=6.37cm]{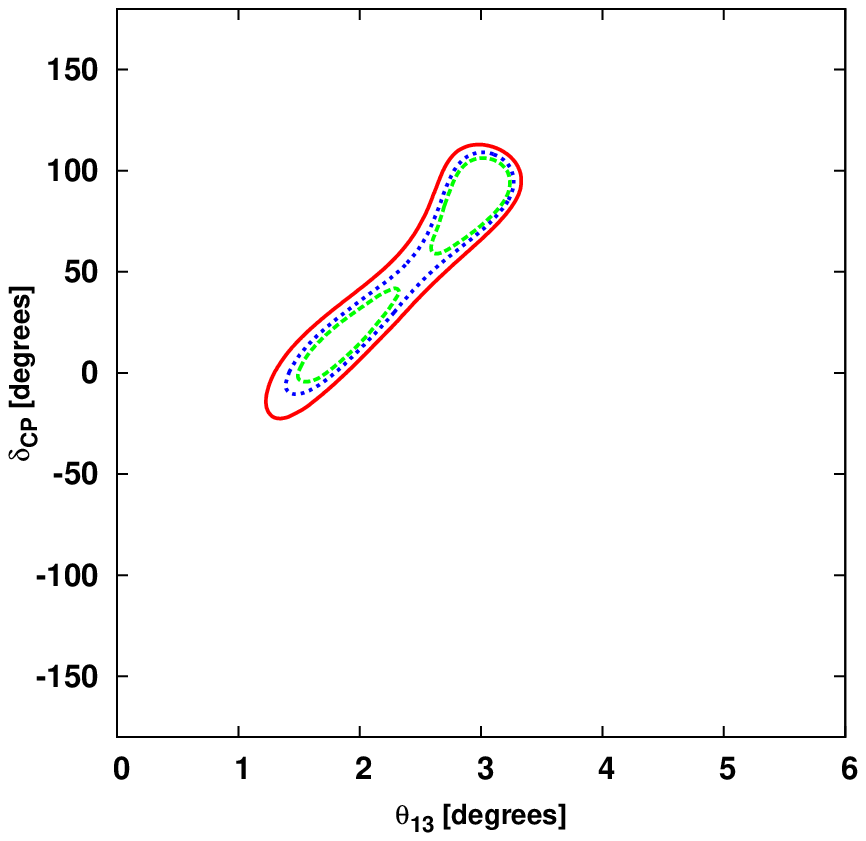}}
\end{center}
\vspace{-5mm}
\caption{90\%, 95\% and 99\% CL contours for the setup III, simulated for 
$\theta_{13}=3^{\circ}$ and $\delta=90^{\circ}$ assuming normal mass ordering and
  $\theta_{23}=45^{\circ}$. The left panel is the contribution of the
  beta-beam, the middle panel is for the electron capture channel with the
  right panel being the combination.}
\label{Fi:CC369} 
\end{figure}

The full power of the combination between the beta-beam spectrum 
and the EC channel is best illustrated in Fig.~\ref{Fi:CC369}. For Setup III, we
give the results with $\theta_{13}=3^{\circ}$ and $\delta=90^{\circ}$ for the 
contribution of the  beta-beam, for that of the EC channel and for the combination.
Each of the two techniques separately suffer from a continuum of
solutions. In fact, the shapes of the allowed regions can be understood
by looking at the form of the oscillation probability.
The power of the combination of the two channels under the same
conditions is in the difference in phase and in amplitude between the
two fake sinusoidal solutions, selecting a narrow allowed region in the
parameter space, much more constrained than the two separate techniques.

       The marked difference between the beta-beam alone and the
combination with the electron capture in this case demonstrate the
importance of data from the high energies.

\subsection{{\bf Disentangling $\theta_{13}$ and $\delta$}}
\begin{figure}
\begin{center}
\hspace{-1.5cm}
\subfigure{\includegraphics[width=8.9cm]{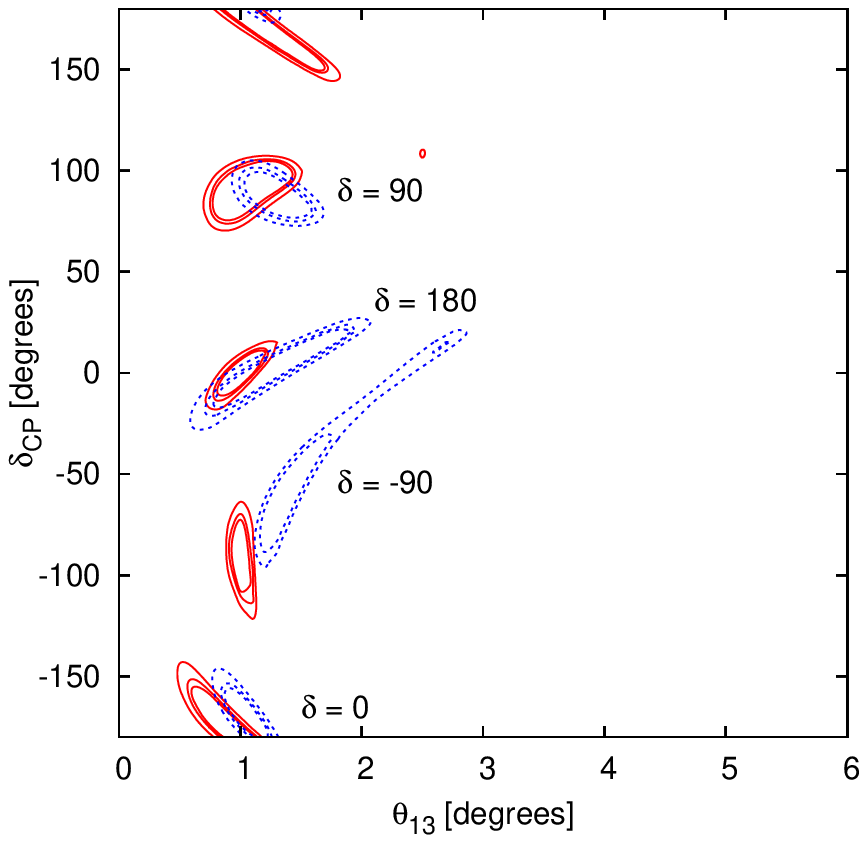}} \hspace{-1.5cm}
\subfigure{\includegraphics[width=8.9cm]{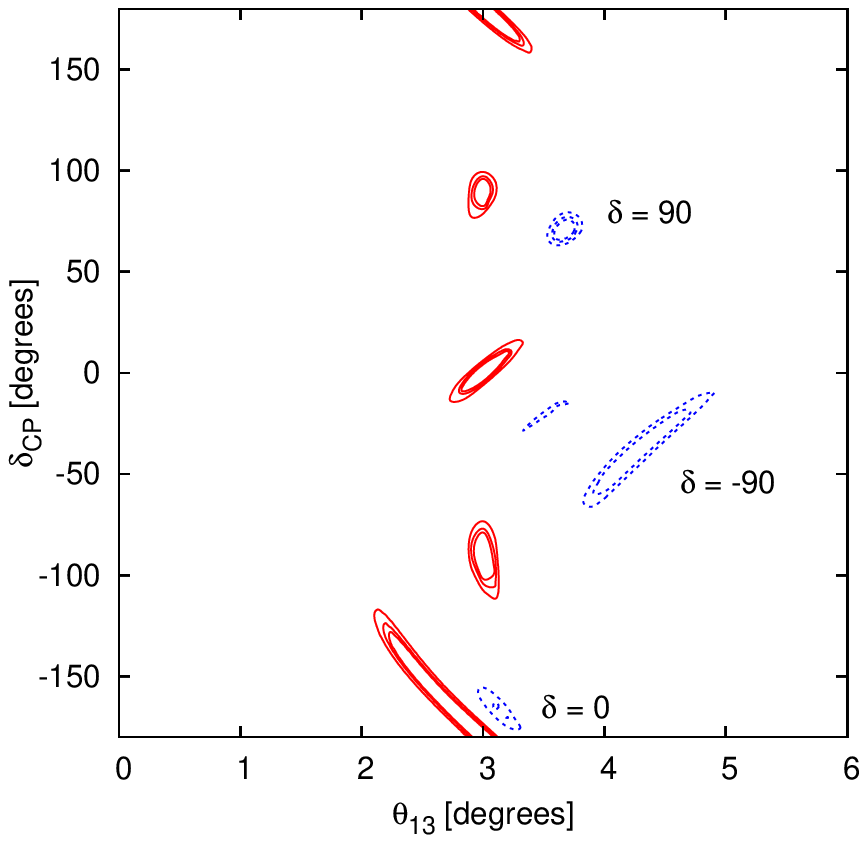}}
\end{center}
\vspace{-2mm}
\caption{90\%, 95\% and 99\% CL contours for setup III-WC with
  solutions from discrete degeneracies included for
  $\theta_{13}=1^{\circ}$ (left panel) and $\theta_{13}=3^{\circ}$
  (right panel) for different values of the CP-phase, $\delta =
  -90^{\circ}, 0^{\circ}, 90^{\circ}, 180^{\circ}$.}
\label{Fi:CC369clone}
\end{figure}
In Fig.~\ref{Fi:CC369clone}, we show the physics reach for setup III-WC,
with the effects of the hierarchy clone solution taken into
account. From a comparison of Figs.~\ref{Fi:CC369}
and~\ref{Fi:CC369clone}, the increase in event rates improves the
results substantially, although not as much as the size factor between the 
two detectors. However, owing to the relatively short
distance, $L=650$~km, the mass ordering can be determined only for
large values of the mixing angle $\theta_{13}$. The
hierarchy degeneracy worsens the ability to measure $\theta_{13}$ and
$\delta$ with good precision, especially for negative true values of
$\delta$. We conclude that a baseline $L=650$~km, at least, is needed to 
disentangle the CP phase.

\section{CP-violation  discovery potential}

We now consider the comparison of setups III-WC and IV-WC for the separation of 
the CP phase $\delta$. Boulby provides a longer baseline $L=1050$~km than Canfranc
 $L=650$~km. This has two contrasting effects on the sensitivity to measure 
CP-violation: i) Sufficient matter effects to resolve the hierarchy degeneracy 
for small values of $\theta_{13}$; ii) It decreases the available statistics.

\begin{figure}
\begin{center}
\hspace{-1.5cm}
\subfigure{\includegraphics[width=8.9cm]{Patall-SIII-WC-Th1.eps}} \hspace{-1.5cm}
\subfigure{\includegraphics[width=8.9cm]{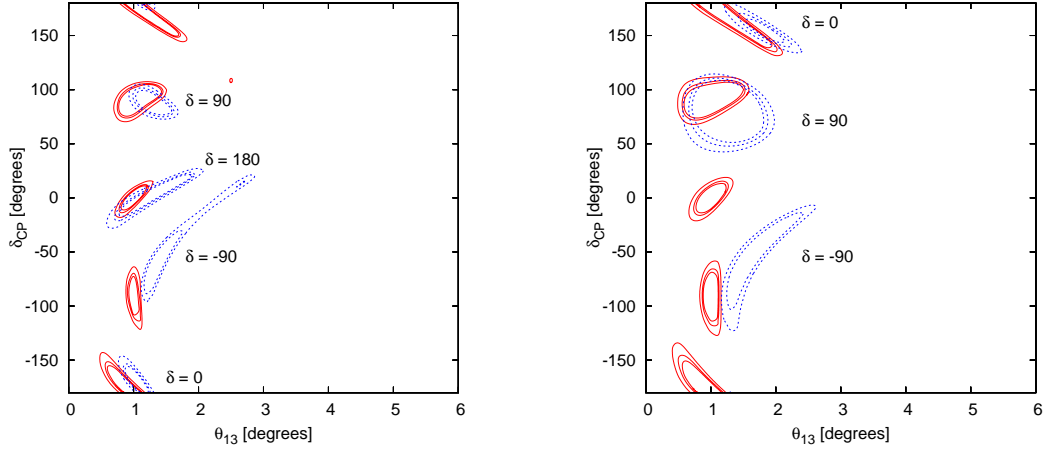}}
\end{center}
\vspace{-2mm}
\caption{90\%, 95\% and 99\% CL contours for setup III-WC (left panel) and IV-WC 
(right panel), with solutions from discrete degeneracies included for
  $\theta_{13}=1^{\circ}$, for different values of the CP-phase, $\delta =
  -90^{\circ}, 0^{\circ}, 90^{\circ}, 180^{\circ}$.}
\label{Fi:CC369clone-WC}
\end{figure}

From Fig.~\ref{Fi:CC369clone-WC}, we see that the smaller count rate results in
 a poorer resolution for the longer baseline. However, the longer baseline allows
 for a good determination of the mass ordering, thus eliminating more degenerate
 solutions.  

\begin{figure}
\begin{center}
\subfigure{\includegraphics[width=7cm]{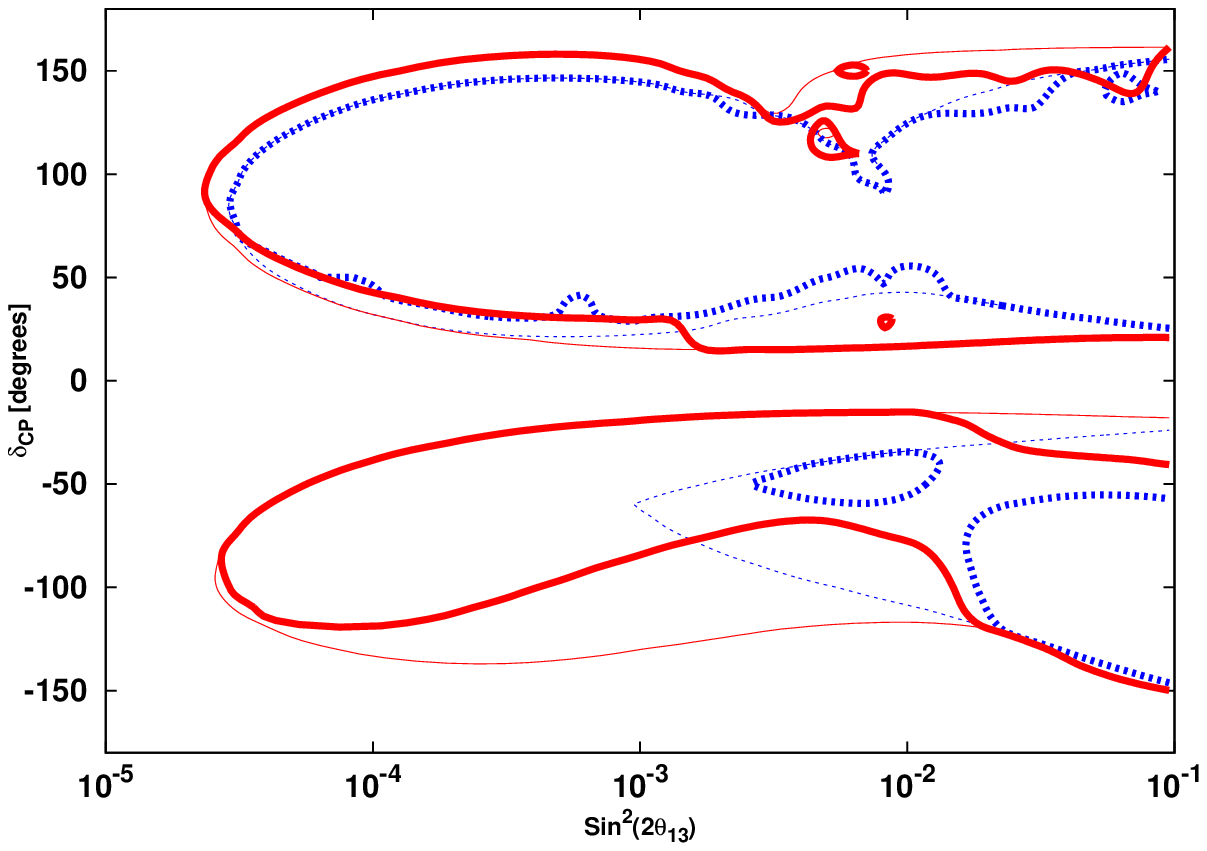}} 
\subfigure{\includegraphics[width=7cm]{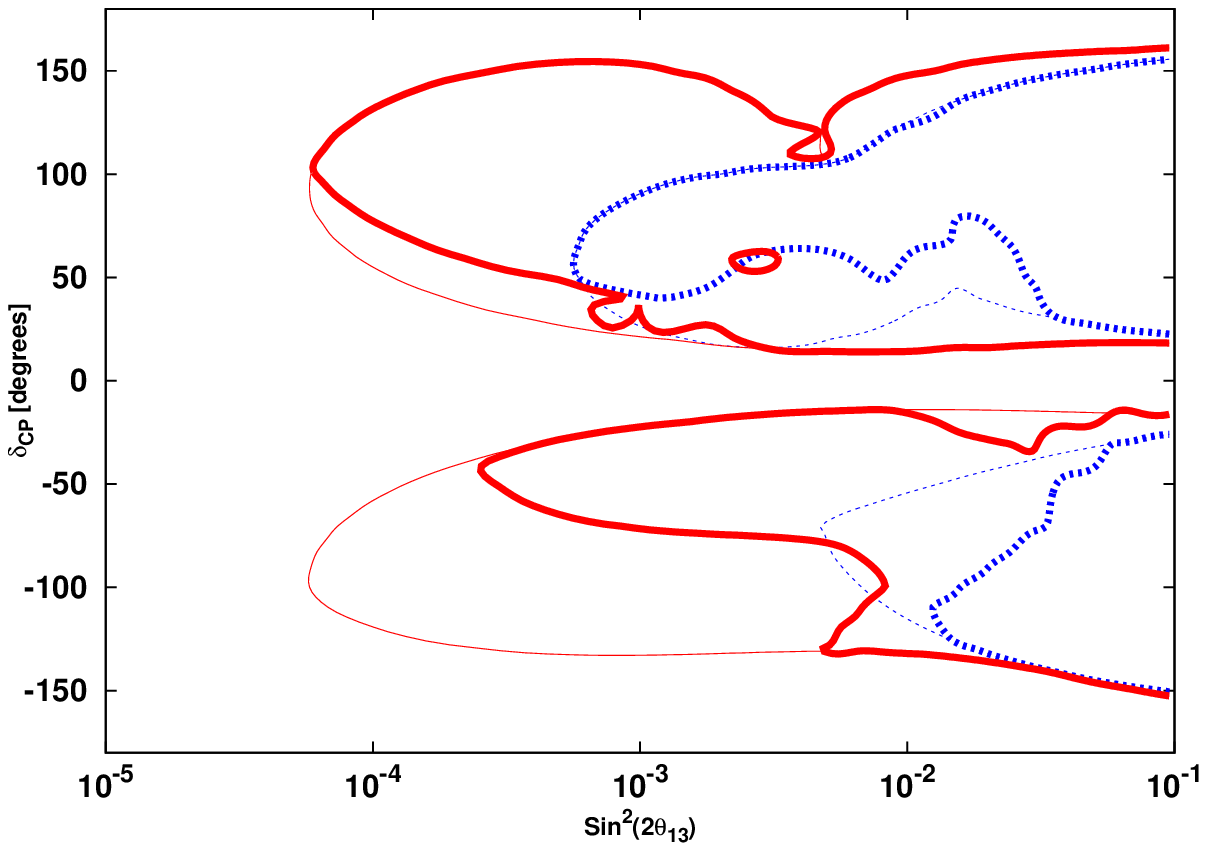}}\\ 
\end{center}
\caption{CP-violation discovery potential at 99\% CL for setup III-WC
  (left panel) and IV-WC (right panel). In each case, we present the
  results for the beta-beam only (blue dotted lines) and the
  combination with the electron capture result (red solid lines), both 
  without (thin lines) and with (thick lines) taking the hierarchy
  degeneracy into account.}
\label{Fi:CPWC}
\end{figure}

In Fig.~\ref{Fi:CPWC} we give the CP-violation discovery potential for these
 two setups. Comparing the two locations of the detector, we notice that the
shorter baseline (CERN-Canfranc) has a slightly (significantly) better
reach for CP-violation at positive (negative) values of $\delta$ than
the longer baseline (CERN-Boulby). The longer option, however, performs
slightly better at negative $\delta$ if the hierarchy is known to be
normal and significantly better if the ordering is not determined.
This is because the longer baseline can identify the neutrino mass
hierarchy for these values of $\theta_{13}$, therefore resolving this
degeneracy.

\newpage

\section{Mass hierarchy determination}
\begin{figure}
\begin{center}
\subfigure{\includegraphics[width=6.8cm]{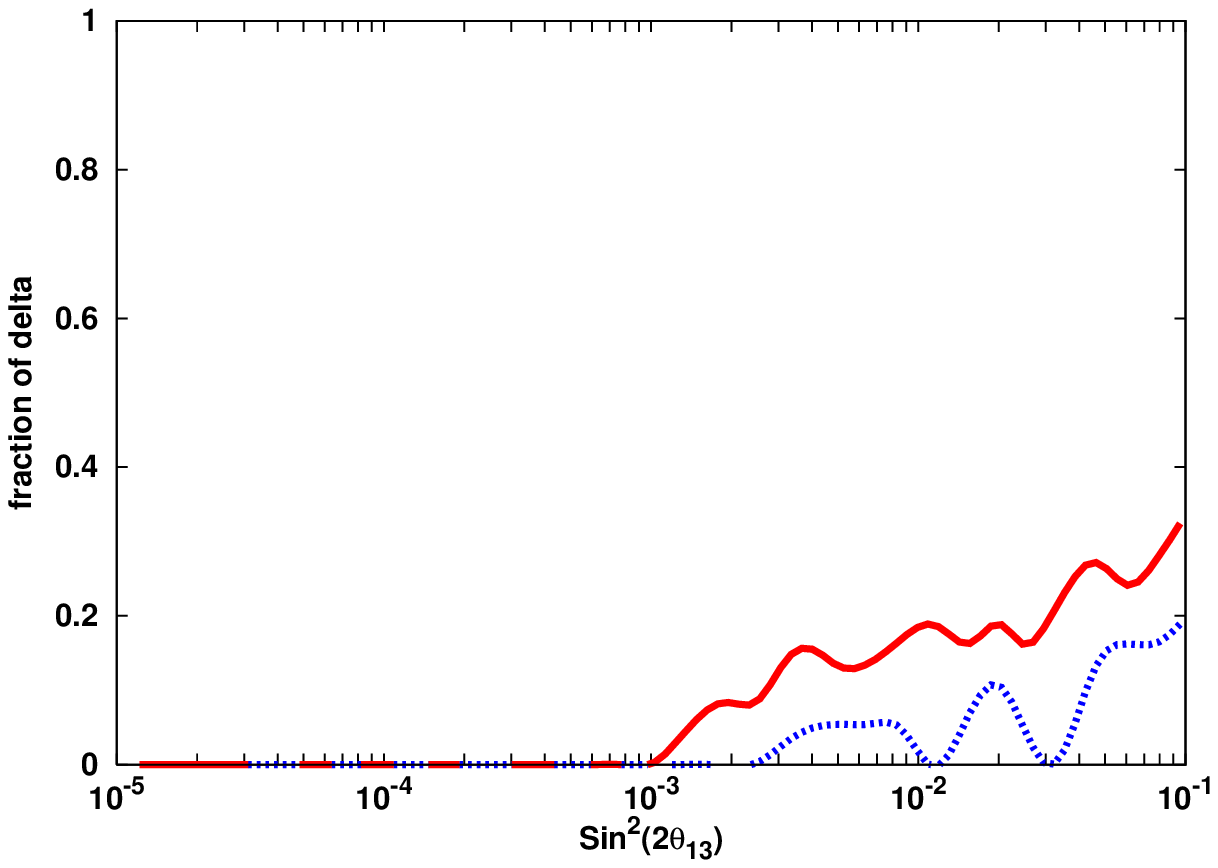}}  \hspace{0.5cm}
\subfigure{\includegraphics[width=6.8cm]{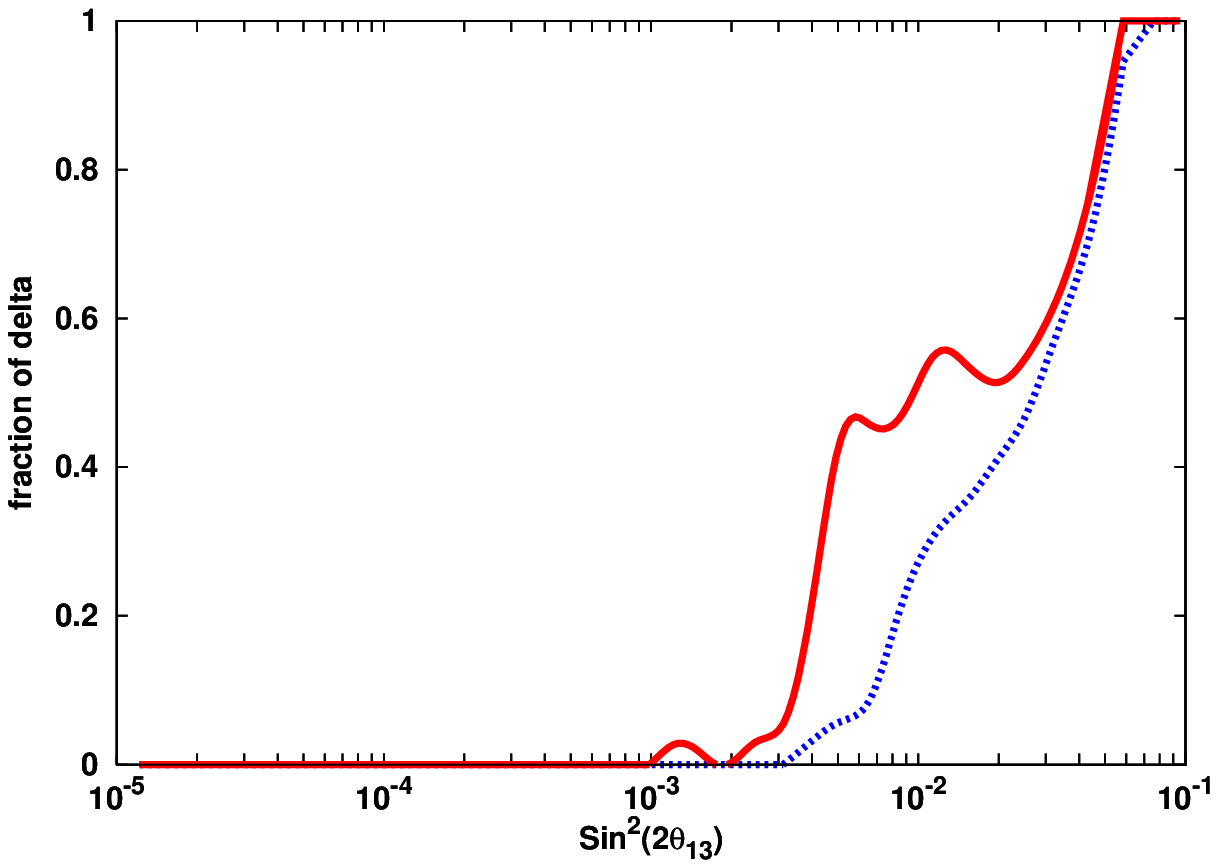}}\\ \vspace{2cm}
\subfigure{\includegraphics[width=6.8cm]{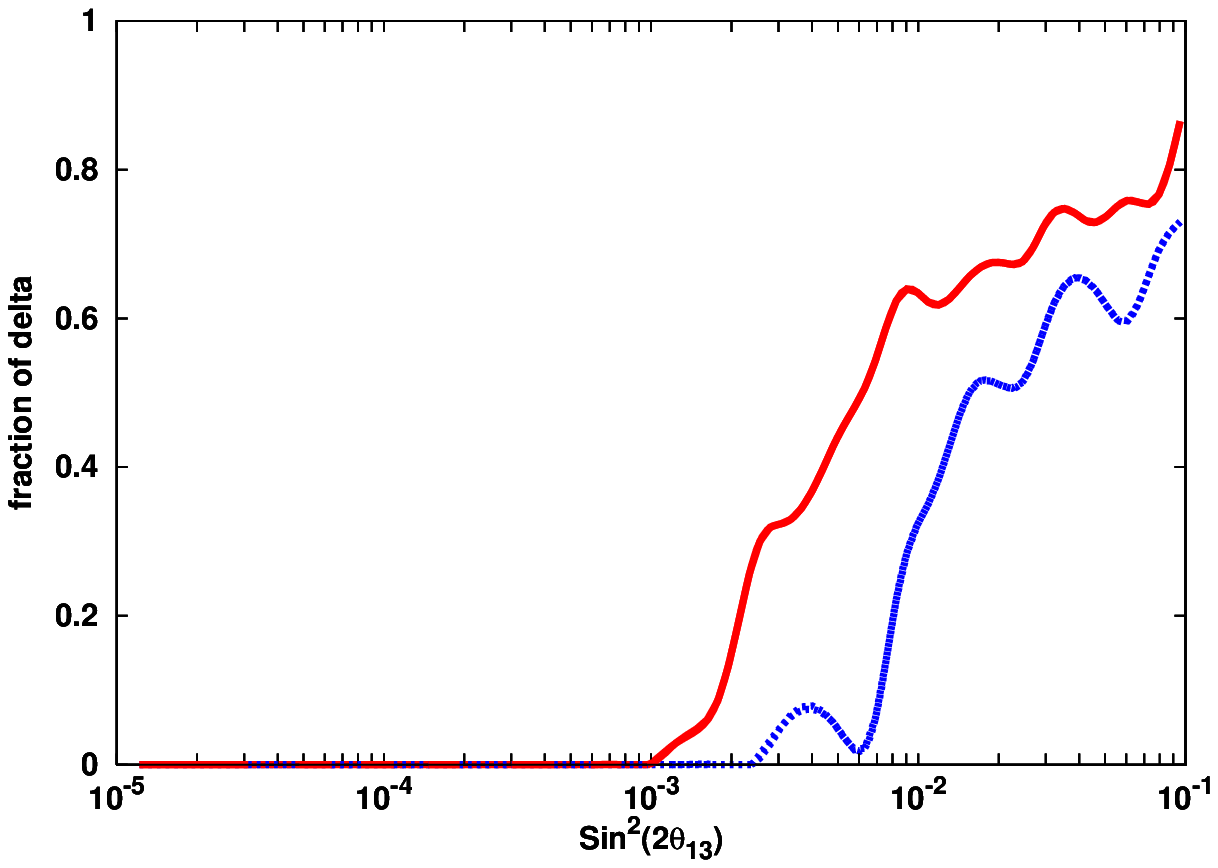}}\\
\end{center}
\caption{Fraction of $\delta$ for which the neutrino mass hierarchy
  can be determined at 99\% CL for setup III-WC (left upper panel) and IV-WC
  (right upper panel), with present priors in the known parameters. The same fraction 
of $\delta$ is presented in the lower panel for setup III-WC with negligible errors in
 the known parameters. In each case, we present the results for the
  beta-beam only (blue dotted lines) and the combination with the
  electron capture result (red solid lines).}
\label{Fi:hierarchyWC}
\end{figure}

In Fig.~\ref{Fi:hierarchyWC}, we present the results for the neutrino
mass hierarchy determination for the setups with a 0.5~Mton
WC detector (setups III-WC and IV-WC). We do not consider the
CERN-Frejus cases; the shorter baseline being unable to distinguish
the type of hierarchy. 

In both cases, the contribution from the beta-beam channel is shown in
blue dashed lines and the result for the combination with the electron
capture channel is shown by the red solid lines. As matter effects are
more important at high energies, we see that the inclusion of the
electron capture flux improves the results.

The CERN-Boulby baseline, with its larger matter effect,
represents a much more promising setup for  the determination of
the mass hierarchy. Its resolution would be possible for all values of $\delta$ for
$\sin^2 2\theta_{13} \simeq \mbox{few} \times 10^{-2}$, even for the present priors in
 the known parameters. The power of having negligible errors in the known parameters is 
presented for setup III-WC associated with the CERN-Canfranc baseline, showing a much
 higher fraction of $\delta$ for which the neutrino mass hierarchy can be determined.

\vspace{1cm}
\section{Conclusions}
We have exploited in this presentation the power of using different neutrino energies, 
in a unique experimental setup, in order to disentangle the CP-odd term in the 
interference between the atmospheric and solar amplitudes for the suppressed 
neutrino oscillation $P(\nu_e \rightarrow \nu_\mu)$. This strategy has been 
implemented by the use of a parent ion with two comparable channels of decay: 
electron capture (EC) and $\beta^+$-decay (BB). The CP phase sensitivity is 
thus obtained by only using neutrinos, thanks to the Energy Dependence of the 
oscillation probability provided by the combination of the two EC and BB channels.

We found that the two separate channels EC and BB have a limited overlap of the 
allowed regions in the $(\theta_{13}, \delta)$ plane, resulting in a good 
resolution of the intrinsic degeneracy. With the SPS upgrade to higher 
energy $(E_p=1000~GeV)$, one gets a better sensitivity to CP-violation  
discovery and measurement, iff accompanied by a longer baseline of
$600$~km at least. This aim is the main focus for the third generation 
neutrino oscillation experiments.

The best $E/L$ for higher sensitivity to the mixing $|U(e3)|$, typically 
the first oscillation maximum, is not the same than that for the CP phase 
determination. Like the phase-shift in a interference pattern, the effect 
of $\delta$ is easier to observe by going to   the energy region of the 
second oscillation. This covering is obtained with the combination of 
the two EC and BB channels for the same ion and the same boost.

The setups III and III-WC, with a baseline $L=650$~km, have larger 
counting rates and a better tuning of the beam to the oscillatory 
patern, resulting in a very good ability for disentangling the two 
parameters $(\theta_{13}, \delta)$. These setups provide the best
 sensitivity for CP-violation with positive values of $\delta$.

 For negative $\delta$'s, the type of hierarchy cannot be still 
resolved for $L=650$~km. In going to setups IV and IV-WC, 
with a baseline $L=1050$~km, the determination of the mass 
hierarchy is better and one obtains a good reach to CP-violation
 for negative $\delta$.

The general conclusion is that the combination of the two EC and BB 
channels from a single decaying ion and a fixed $\gamma$-boost 
achieves remarkable results for the discovery of CP-violation 
and the measurement of the CP-phase. This is a virtue of the 
different energy dependence of the CP-odd term and the CP-even 
terms in the oscillation probability, and there is no need of
 performing separate experiments, with different systematics 
and counting rates, for neutrinos and antineutrinos.

\section{Acknowledgements}
It is a pleasure to acknowledge Milla for the magnificent event 
organized in Venize. We would like to thank  many colleagues
 for discussions and clarifications and, particularly, to our 
collaborators J.~Burguet-Castell,  M.~Lindroos, C.~Orme, 
S.~ Palomares-Ruiz and S.~Pascoli. This research has been 
funded by the Spanish MICINN Grant FPA2008-02878 and by the 
Generalitat Valenciana Grant PROMETEO 2088/004.

 \end{document}